\documentclass[submission, Proceedings]{SciPost}

\hypersetup{
    colorlinks,
    linkcolor={red!50!black},
    citecolor={blue!50!black},
    urlcolor={blue!80!black}
}

\usepackage{lineno}
\usepackage[bitstream-charter]{mathdesign}
\usepackage{caption}
\usepackage{subcaption}
\usepackage{upgreek}

\DeclareSymbolFont{usualmathcal}{OMS}{cmsy}{m}{n}
\DeclareSymbolFontAlphabet{\mathcal}{usualmathcal}

\begin{document}

\begin{center}{\Large \textbf{
Recent results on Central Exclusive Production\\with the STAR detector\\
}}\end{center}

% TODO: write the author list here. Use initials + surname format.
% Separate subsequent authors by a comma, omit comma at the end of the list.
% Mark the corresponding author with a superscript *.
\begin{center}
Rafał Sikora\textsuperscript{$\star$} for the STAR Collaboration
\end{center}

\begin{center}
Faculty of Physics and Applied Computer Science, AGH University of Science and Technology, 30 Mickiewicza Ave., 30-059 Kraków, Poland
\\
* rafal.sikora@agh.edu.pl
\end{center}

\begin{center}
\today
\end{center}

% For convenience during refereeing (optional),
% you can turn on line numbers by uncommenting the next line:
%\linenumbers

\definecolor{palegray}{gray}{0.95}
\begin{center}
\colorbox{palegray}{
  \begin{tabular}{rr}
  \begin{minipage}{0.1\textwidth}
    \includegraphics[width=22mm]{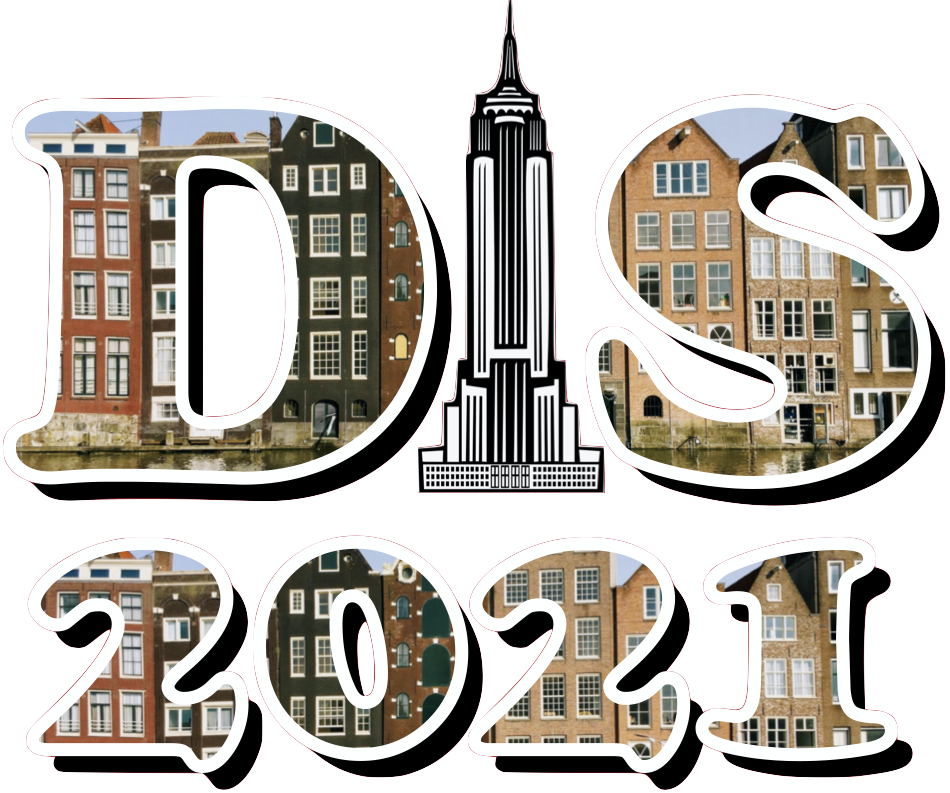}
  \end{minipage}
  &
  \begin{minipage}{0.75\textwidth}
    \begin{center}
    {\it Proceedings for the XXVIII International Workshop\\ on Deep-Inelastic Scattering and
Related Subjects,}\\
    {\it Stony Brook University, New York, USA, 12-16 April 2021} \\
    \doi{10.21468/SciPostPhysProc.?}\\
    \end{center}
  \end{minipage}
\end{tabular}
}
\end{center}

\section*{Abstract}
{\bf
% TODO: write your abstract here.
We present results on the Central Exclusive Production of charged particle pairs $h^{+}h^{-}$ ($h = \pi, K, p$), $pp\to p^\prime+h^{+}h^{-}+p^\prime$, obtained with the STAR experiment at RHIC in proton-proton collisions at a center-of-mass energy of $\sqrt{s} = 200$~GeV. All final-state particles were reconstructed, including forward-scattered protons detected in the Roman Pot system. As a result, the Double Pomeron Exchange (DPE) events were selected and the non-exclusive backgrounds were efficiently rejected.

Differential fiducial cross sections were measured as functions of observables related to the central hadronic final state and to the forward-scattered protons. The measured cross sections were compared to phenomenological model predictions based on the DPE.% Structures observed in the mass spectra of $\pi^{+}\pi^{-}$ and $K^{+}K^{-}$ pairs were found consistent with the DPE model, while angular distributions of pions suggested a dominant spin-0 contribution to $\pi^{+}\pi^{-}$ production. 

%For $\pi^+\pi^-$ production, the fiducial cross section was extrapolated to the Lorentz-invariant region and was successfully modeled assuming the continuum production and at least three resonances, the $f_0(980)$, $f_2(1270)$, and $f_0(1500)$, with a possible small contribution from the $f_0(1370)$.  
%Fits to the extrapolated differential cross section as a~function of squared four-momentum transfers in proton vertices enabled extraction of the exponential slope parameters in several bins of the invariant mass of $\pi^+\pi^-$ pairs. These parameters are sensitive to the size of the interaction region.

We also present preliminary results on the measurement of the same physics process at higher collision energy $\sqrt{s} = 510$~GeV. The data demonstrate features similar to those observed at $\sqrt{s} = 200$~GeV.
}

%\vspace{10pt}
%\noindent\rule{\textwidth}{1pt}
%\tableofcontents\thispagestyle{fancy}
%\noindent\rule{\textwidth}{1pt}
%\vspace{10pt}

\section{Introduction}
\label{sec:intro}
The Central Exclusive Production (CEP) process in proton-proton collisions, $p+p\rightarrow p'+X+p'$, is experimentally recognized when all particles of the centrally-produced neutral state $X$ (here denoting a pair of opposite-charge hadrons, $h^+h^-$) are well separated in the rapidity space from the intact scattered incoming protons ($p'$). At sufficiently high center-of-mass energies the process occurs mainly through the Double Pomeron Exchange (DPE) mechanism, which is considered suitable for the production of the gluon bound states (glueballs). In this process the hadron pair continuum and regular resonances are produced as well, all interfering with each other. Additional soft exchanges are also possible between the particles involved in the process, spoiling the exclusivity of the reaction (so-called absorption and rescattering effects). Theoretical complexity of this topologically-simple process makes it a relevant subject for experimental and theoretical studies~\cite{LS,Durham,MBR}. 

\section{Results}
\begin{figure}[b!]
	\centering
	\parbox{0.485\textwidth}{%
		\centering
		\begin{subfigure}[b]{\linewidth}{
				\subcaptionbox{\vspace*{-0pt}\label{fig:tSum}}{\includegraphics[width=\linewidth]{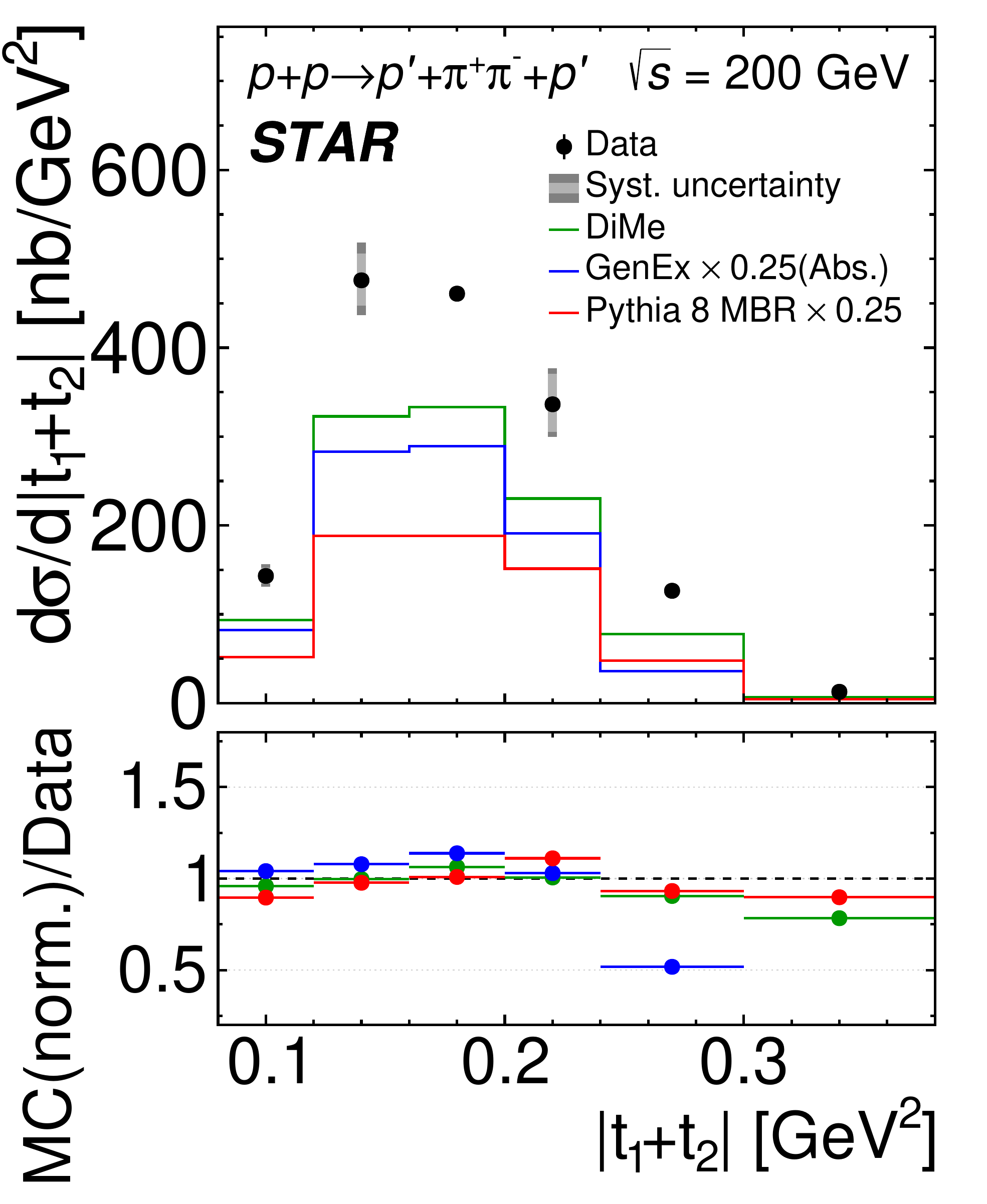}\vspace{-10pt}}}\vspace{-7pt}
		\end{subfigure}
	}%%
	\quad%
	\parbox{0.485\textwidth}{
		\centering
		\begin{subfigure}[b]{\linewidth}{
				\subcaptionbox{\vspace*{-0pt}\label{fig:deltaPhi}}{\includegraphics[width=\linewidth]{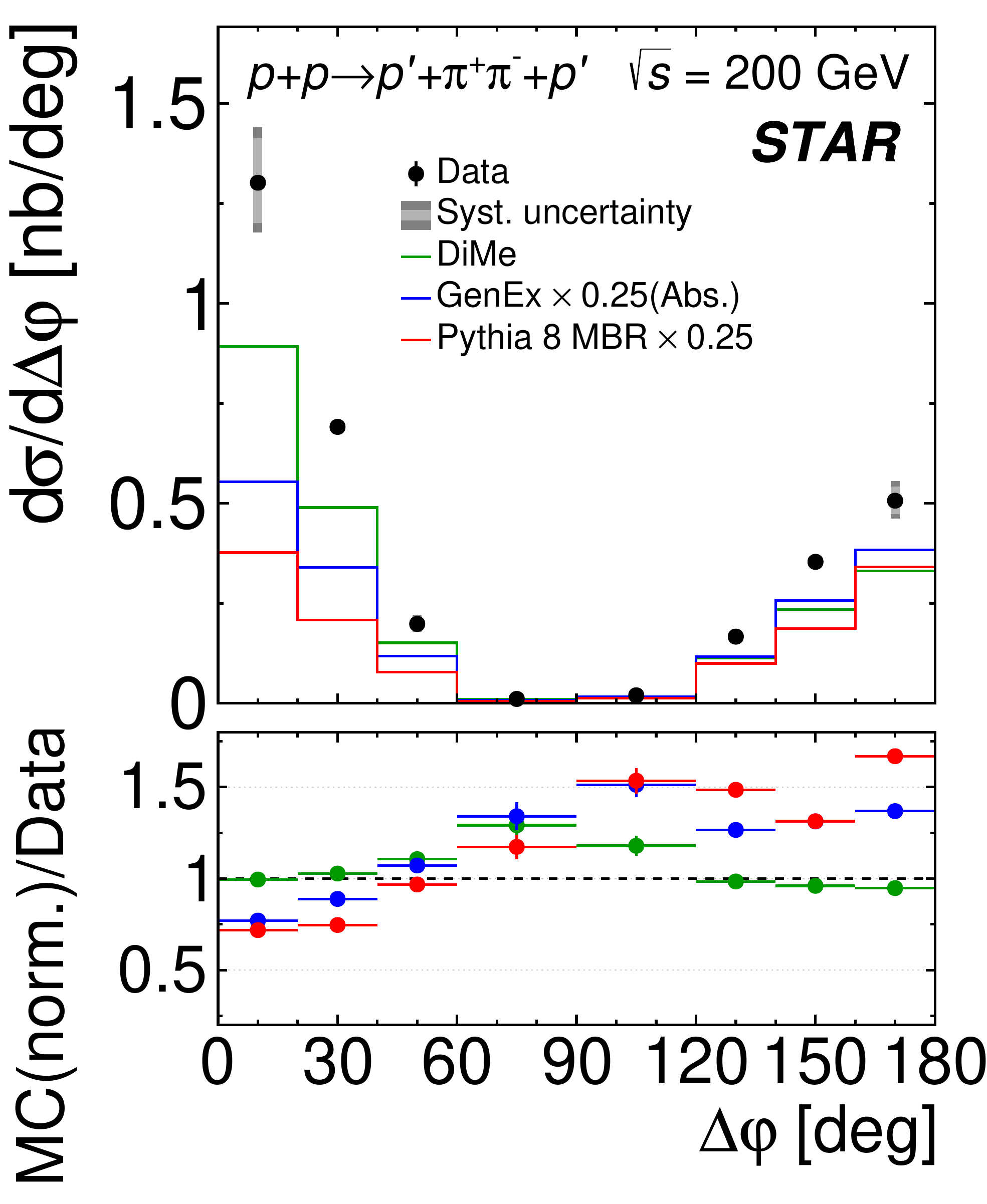}\vspace{-10pt}}}\vspace{-7pt}
		\end{subfigure}
	}%
	\caption{Differential fiducial cross sections for CEP of $\pi^+\pi^-$ pairs as a function of (\subref{fig:tSum}) the sum of the squares of the four-momentum transfers in the proton vertices, and (\subref{fig:deltaPhi}) the difference of azimuthal angles of the forward-scattered protons~\cite{cepSTAR}. Data are shown as solid points with error bars representing the statistical uncertainties. The typical systematic uncertainties are shown as grey boxes for only a few data points as they are almost fully correlated between neighbouring bins. Predictions from Monte Carlo models GenEx~\cite{LS}, DiMe~\cite{Durham} and MBR~\cite{MBR} are shown as histograms.}
	\label{fig:deltaPhi_tSum}
\end{figure}

The STAR experiment~\cite{STAR} at RHIC has performed a high-statistics measurement of the CEP process in proton-proton collisions at the center-of-mass energy $\sqrt{s}=200$~GeV~\cite{cepSTAR}. Figures~\ref{fig:tSum} and~\ref{fig:deltaPhi} show the differential fiducial cross sections for CEP of $\pi^+\pi^-$ pairs as a function of the sum of the squared four-momentum transfers, $|t_1+t_2|$, and the azimuthal separation, $\Delta\upvarphi$, of the forward-scattered protons, respectively. These physics observables, whose reconstruction was possible thanks to dedicated forward proton detectors mounted in the Roman Pot vessels, are particularly sensitive to the absorption effects.

\begin{figure}%[b!]
	\centering
	\parbox{0.485\textwidth}{
		\centering
		\begin{subfigure}[b]{\linewidth}{
				\subcaptionbox{\vspace*{-0pt}\label{fig:invMass_2pi_STAR_DeltaPhiLess90}}{\includegraphics[width=\linewidth]{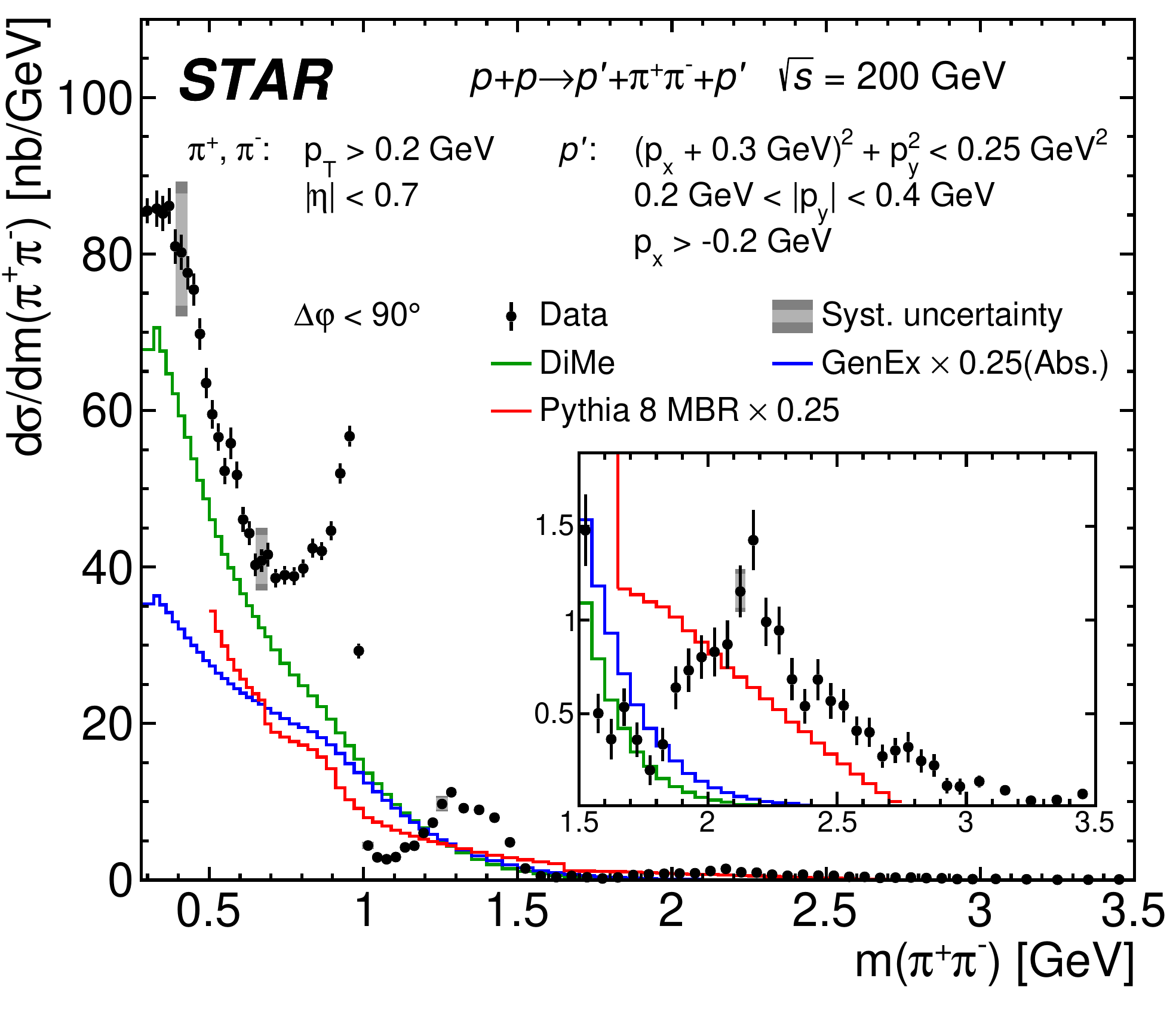}\vspace{-10pt}}}\vspace{-7pt}
		\end{subfigure}
	}%
	\quad%
	\parbox{0.485\textwidth}{%
		\centering
		\begin{subfigure}[b]{\linewidth}{
				\subcaptionbox{\vspace*{-0pt}\label{fig:invMass_2pi_STAR_DeltaPhiMore90}}{\includegraphics[width=\linewidth]{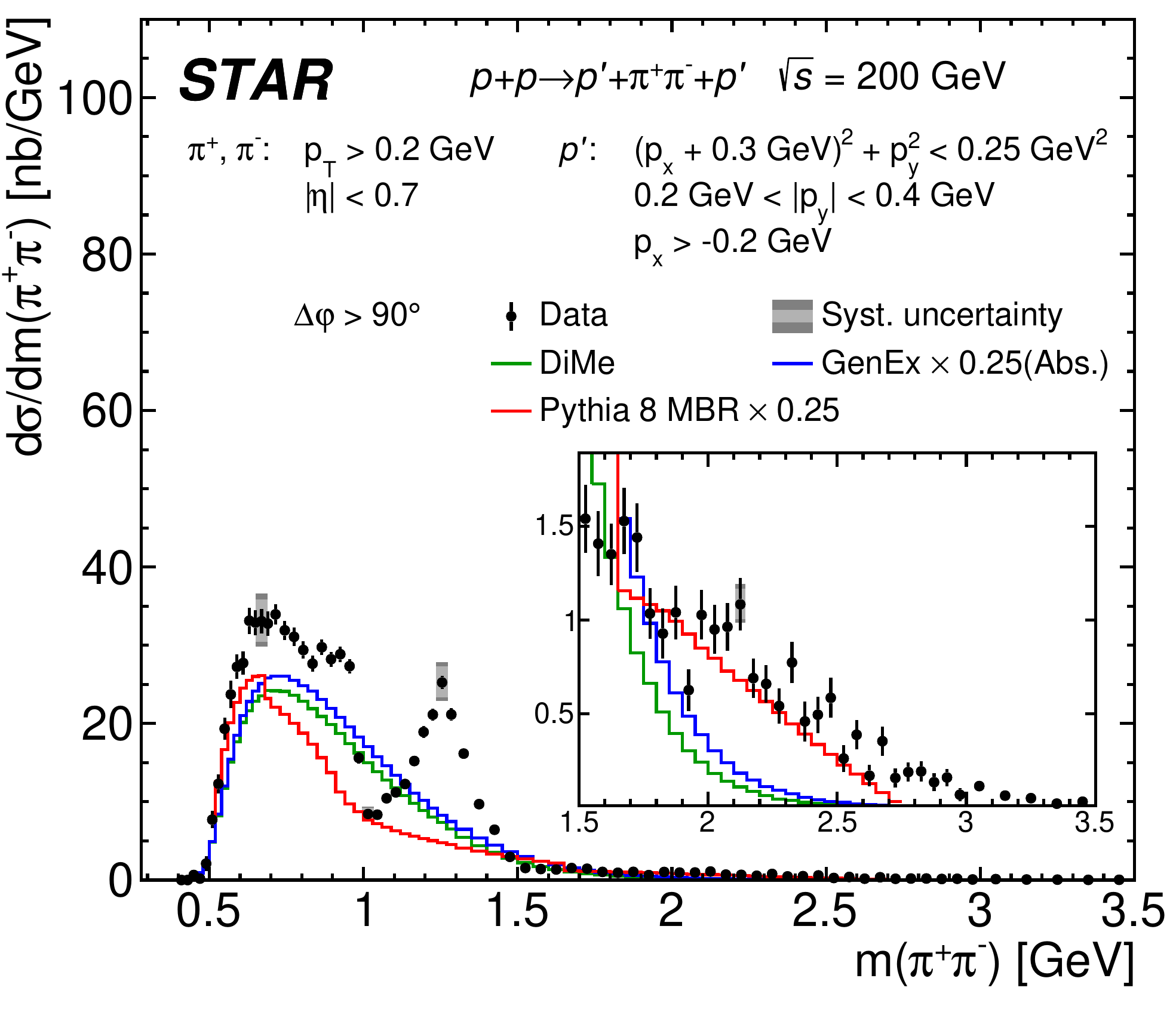}\vspace{-10pt}}}\vspace{-7pt}
		\end{subfigure}
	}
	\caption[Differential cross sections for CEP of $\pi^+\pi^-$ pairs as a function of the invariant mass of the pair in two $\Delta\upvarphi$ regions: $\Delta\upvarphi<90^\circ$ and $\Delta\upvarphi>90$ degree, measured in the fiducial region explained on the plots.]{Differential cross sections for CEP of $\pi^+\pi^-$ pairs as a function of the invariant mass of the pair in two $\Delta\upvarphi$ regions: (\subref{fig:invMass_2pi_STAR_DeltaPhiLess90}) $\Delta\upvarphi<90^\circ$ and (\subref{fig:invMass_2pi_STAR_DeltaPhiMore90}) $\Delta\upvarphi>90^\circ$, measured in the fiducial region explained on the plots~\cite{cepSTAR}. Data are shown as solid points with error bars representing the statistical uncertainties. The typical systematic uncertainties are shown as grey boxes for only a few data points as they are almost fully correlated between neighbouring bins. Predictions from Monte Carlo models GenEx, DiMe and MBR are shown as histograms.}
	\label{fig:invMass_2pi_STAR_DeltaPhi}
\end{figure}

\begin{figure}[b!]
	\centering
	\parbox{0.485\textwidth}{
		\centering
		\begin{subfigure}[b]{\linewidth}{
				\subcaptionbox{\vspace*{-0pt}\label{fig:invMass_kk}}{\includegraphics[width=\linewidth]{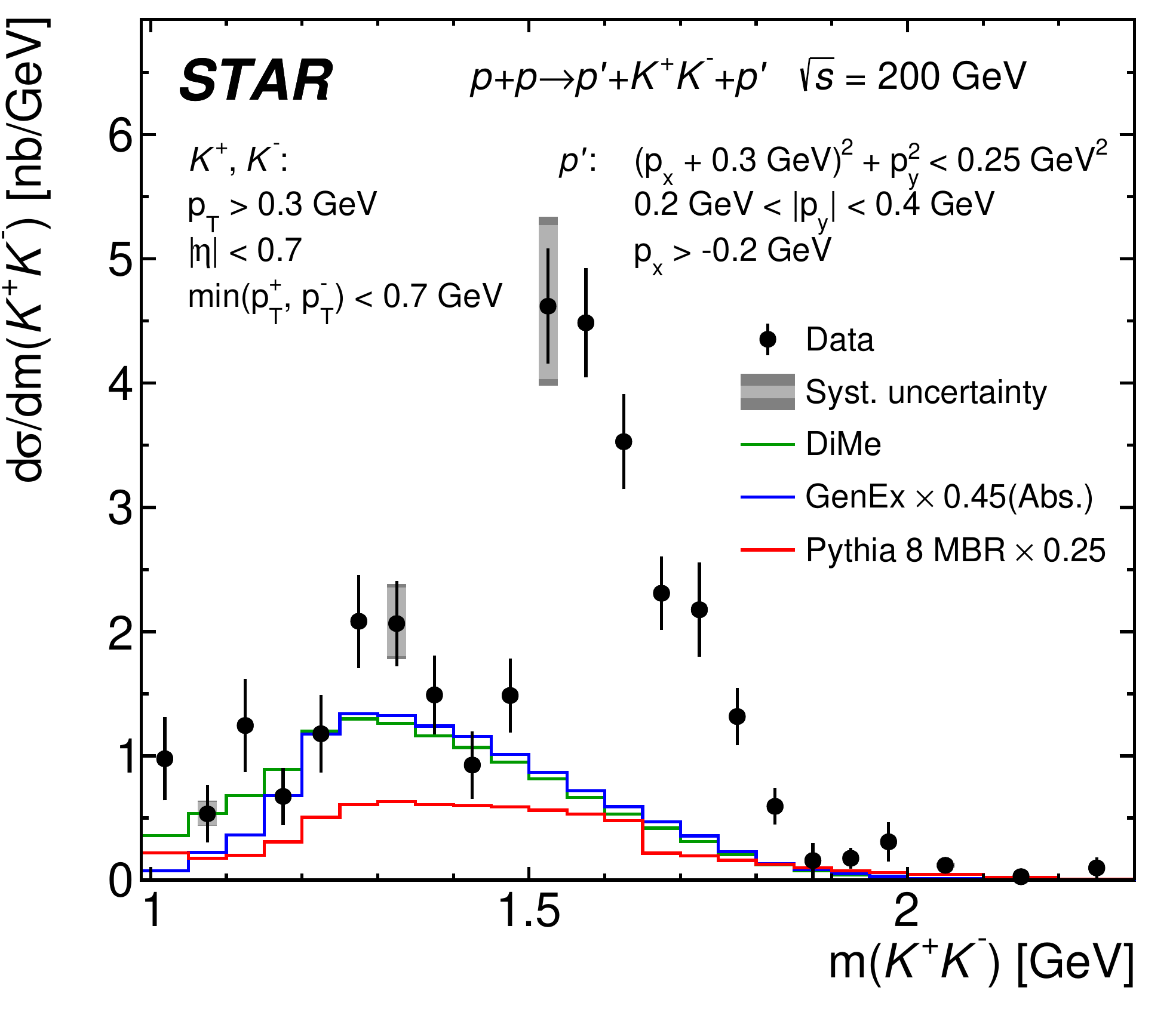}\vspace{-10pt}}}\vspace{-7pt}
		\end{subfigure}
	}%
	\quad%
	\parbox{0.485\textwidth}{%
		\centering
		\begin{subfigure}[b]{\linewidth}{
				\subcaptionbox{\vspace*{-0pt}\label{fig:invMass_ppbar}}{\includegraphics[width=\linewidth]{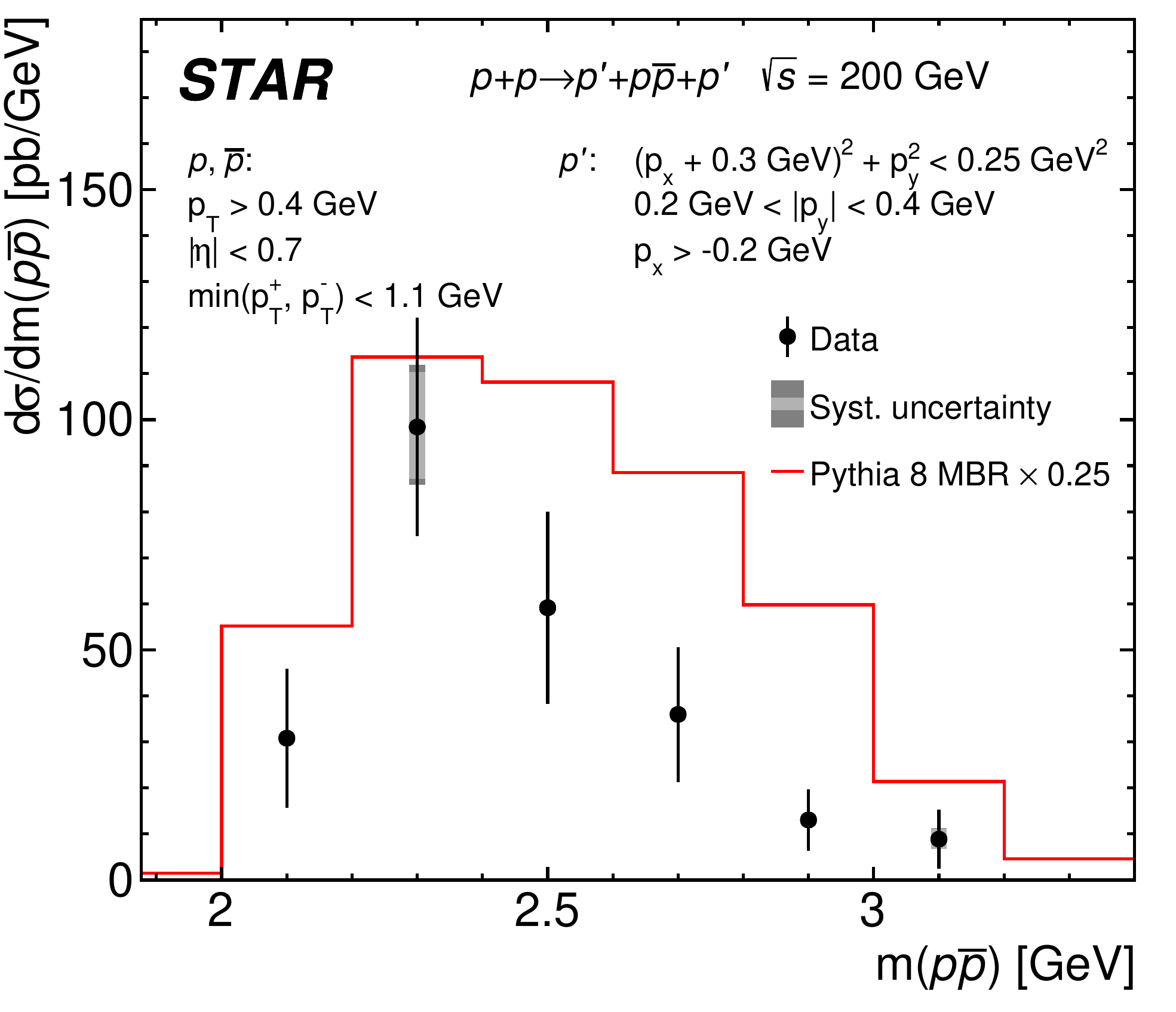}\vspace{-10pt}}}\vspace{-7pt}
		\end{subfigure}
	}
	\caption{Differential cross sections for CEP of (\subref{fig:invMass_kk}) $K^+K^-$ and (\subref{fig:invMass_ppbar}) $p\bar{p}$ pairs as a function of the invariant mass of the pair, measured in the fiducial region explained on the plots~\cite{cepSTAR}. Data are shown as solid points with error bars representing the statistical uncertainties. The typical systematic uncertainties are shown as grey boxes for only a few data points as they are almost fully correlated between neighbouring bins. Predictions from Monte Carlo models GenEx, DiMe and MBR are shown as histograms.}
	\label{fig:invMass_kk_ppbar}
\end{figure}

Properties of the central hadron pairs were studied with respect to the angular configuration of the forward-scattered protons, which provides sensitivity to dynamical effects of the Pomeron exchange. Figure~\ref{fig:invMass_2pi_STAR_DeltaPhi} shows the invariant mass of the $\pi^+\pi^-$ pairs in two characteristic ranges of $\Delta\upvarphi$ angle. Comparison of the spectra reveals significant differences in the magnitudes of the structures attributed to $f_0(980)$, $f_2(1270)$, and potential resonance around 2.2~GeV, in the two regions of $\Delta\upvarphi$ angle. Similar effects have been observed in the CEP of $K^+K^-$ and $p\bar{p}$ pairs, for which the differential fiducial cross sections as a function of the pair invariant mass are shown in Fig.~\ref{fig:invMass_kk_ppbar}.

\begin{figure}[t!] 
	\centering
	\includegraphics[width=\textwidth,page=1]{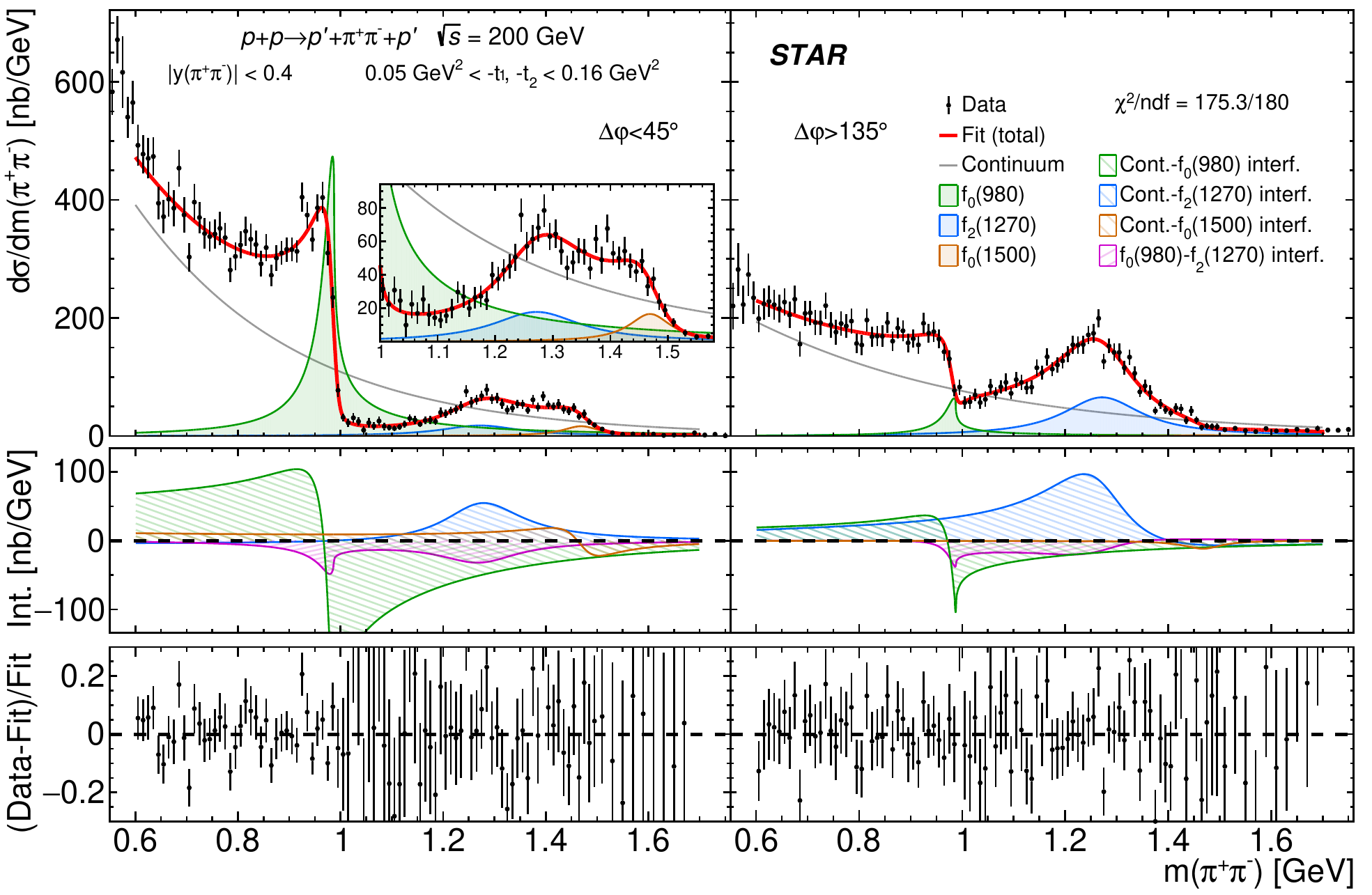}
	\caption[Differential cross section $d\sigma/dm(\pi^{+}\pi^{-})$ extrapolated from the fiducial region to $|y(\pi^{+}\pi^{-})|<0.4$ and $0.05~\text{GeV}^{2} < -t_{1}, -t_{2} < 0.16~\text{GeV}^{2}$]{Differential cross section $d\sigma/dm(\pi^{+}\pi^{-})$ extrapolated from the fiducial region to the Lorentz-invariant phase space given by the central-state rapidity, $|y(\pi^{+}\pi^{-})|<0.4$, and squared four-momentum transferred in forward proton vertices, $0.05\,\text{GeV}^{2} < -t_1, -t_2 < 0.16\,\text{GeV}^{2}$~\cite{cepSTAR}. The left and right panels show the cross sections for $\Delta\upvarphi<45^\circ$ and $\Delta\upvarphi>135^\circ$, respectively. The data are shown as black points with error bars representing statistical uncertainties. The result of the fit is drawn with a solid red line. The squared amplitudes for the continuum and resonance production are drawn with lines of different colors, as explained in the legend. The most significant interference terms are plotted in the middle panels, while the relative differences between each data point and the fitted model are shown in the bottom panels.}
	\label{fig:invMassFit}
\end{figure}

Differential cross section as a function of the invariant mass of exclusively produced $\pi^+\pi^-$ pairs measured in the fiducial region was extrapolated to the Lorentz-invariant phase space volume (Fig.~\ref{fig:invMassFit}). This enabled fitting of the spectrum and extraction of contributions from resonant states and a continuum. The resonance consistent with $f_0(1500)$ was observed in the region $\Delta\upvarphi<45^\circ$, which might suggest that this state contains a purely gluonic component. Extracted properties of the resonances, including their cross sections, should help constrain DPE models.

\begin{figure}[t!]
	\centering
	\parbox{0.485\textwidth}{
		\centering
		\begin{subfigure}[b]{\linewidth}{
				\subcaptionbox{\vspace*{-0pt}\label{fig:invMass_pipi_510}}{\includegraphics[width=\linewidth]{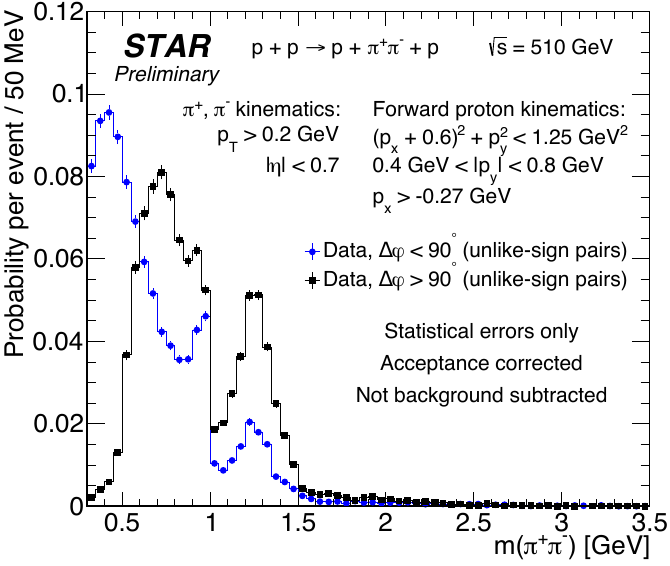}\vspace{-10pt}}}\vspace{-7pt}
		\end{subfigure}
	}%
	\quad%
	\parbox{0.485\textwidth}{%
		\centering
		\begin{subfigure}[b]{\linewidth}{
				\subcaptionbox{\vspace*{-0pt}\label{fig:invMass_kk_510}}{\includegraphics[width=\linewidth]{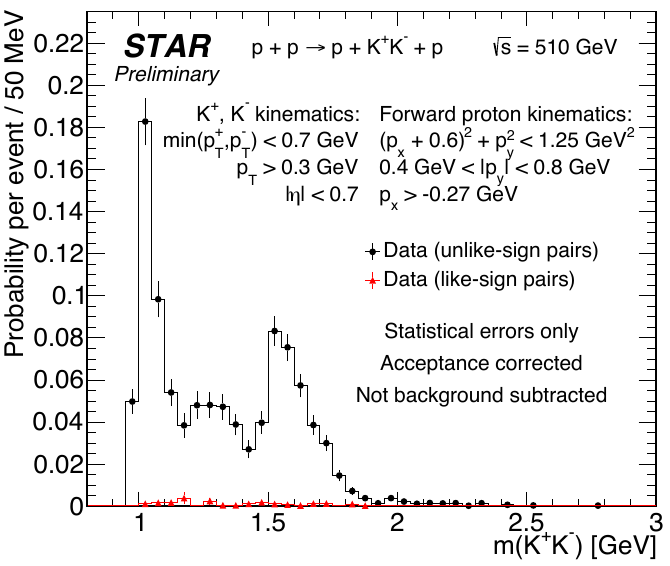}\vspace{-10pt}}}\vspace{-7pt}
		\end{subfigure}
	}
	\caption{Uncorrected preliminary distributions of the invariant mass of (\subref{fig:invMass_pipi_510}) $\pi^+\pi^-$ and (\subref{fig:invMass_kk_510}) $K^+K^-$ pairs, produced in the CEP process at $\sqrt{s}=510$~GeV, measured in the fiducial region explained on the plots. Error bars represent the statistical uncertainties. Figure~(\subref{fig:invMass_pipi_510}) shows data separately for $\Delta\upvarphi<90^\circ$ (blue) and $\Delta\upvarphi>90^\circ$ (black).}
	\label{fig:invMass_kk_ppbar_510}
\end{figure}

The STAR experiment collected CEP data also at a higher center-of-mass energy of 510~GeV. Studies similar to those performed at $\sqrt{s}=200$~GeV are ongoing, which will enable direct comparisons of the CEP process at different $\sqrt{s}$. Figure~\ref{fig:invMass_kk_ppbar_510} shows the preliminary invariant mass spectra of exclusively produced $\pi^{+}\pi^{-}$ and $K^+K^-$ pairs at $\sqrt{s}=510$~GeV. These preliminary data are qualitatively compatible with the published results at $\sqrt{s}$=200~GeV. However some differences are visible e.g. the magnitude of the peak at the invariant mass of 1~GeV for $K^+K^-$ (Fig.~\ref{fig:invMass_kk_510}), which require further investigations.

\section{Conclusions}
Recent results on CEP of charged hadron pairs: $\pi^+\pi^-$, $K^+K^-$ and $p\bar{p}$, in proton-proton collisions at $\sqrt{s}=200$~GeV (published results) and $\sqrt{s}=510$~GeV (preliminary results) obtained by the STAR experiment have been presented. These are currently the highest center-of-mass energies, at which this process is measured with the detection of the forward-scattered protons. The obtained results are expected to help in understanding the Pomeron structure and developing phenomenological models of the DPE process.

\section*{Acknowledgements}
This work was partly supported by the National Science Centre of Poland under grant number UMO-2018/30/M/ST2/00395.

%\bibliographystyle{SciPostbibstyle} % Include this style file here only if you are not using our template
\bibliography{references.bib}

\begin{thebibliography}{1}
\providecommand{\url}[1]{\texttt{#1}}
\providecommand{\urlprefix}{URL }
\expandafter\ifx\csname urlstyle\endcsname\relax
  \providecommand{\doi}[1]{doi:\discretionary{}{}{}#1}\else
  \providecommand{\doi}{doi:\discretionary{}{}{}\begingroup
  \urlstyle{rm}\Url}\fi
\providecommand{\eprint}[2][]{\url{#2}}

\bibitem{LS}
P.~Lebiedowicz and A.~Szczurek,
\newblock \emph{{Exclusive $pp \to pp \pi^{+}\pi^{-}$ reaction: From the
  threshold to LHC}},
\newblock Phys. Rev. D \textbf{81}, 036003 (2010),
\newblock \doi{10.1103/PhysRevD.81.036003},
\newblock \eprint[arXiv]{0912.0190}.

\bibitem{Durham}
L.~Harland-Lang, V.~Khoze and M.~Ryskin,
\newblock \emph{{Modelling exclusive meson pair production at hadron
  colliders}},
\newblock Eur. Phys. J. C \textbf{74}, 2848 (2014),
\newblock \doi{10.1140/epjc/s10052-014-2848-9},
\newblock \eprint[arXiv]{1312.4553}.

\bibitem{MBR}
R.~Ciesielski and K.~Goulianos,
\newblock \emph{{MBR Monte Carlo Simulation in PYTHIA8}},
\newblock PoS \textbf{ICHEP2012}, 301 (2013),
\newblock \doi{10.22323/1.174.0301},
\newblock \eprint[arXiv]{1205.1446}.

\bibitem{STAR}
K.~Ackermann \emph{et~al.},
\newblock \emph{{STAR detector overview}},
\newblock Nucl. Instrum. Meth. A \textbf{499}, 624 (2003),
\newblock \doi{10.1016/S0168-9002(02)01960-5}.

\bibitem{cepSTAR}
J.~Adam \emph{et~al.},
\newblock \emph{{Measurement of the central exclusive production of charged
  particle pairs in proton-proton collisions at $\sqrt{s} = 200$ GeV with the
  STAR detector at RHIC}},
\newblock JHEP \textbf{07}(07), 178 (2020),
\newblock \doi{10.1007/JHEP07(2020)178},
\newblock \eprint[arXiv]{2004.11078}.

\end{thebibliography}

\nolinenumbers

\end{document}